\documentclass[aps,pre,eqsecnum,preprint,tightenlines,nofootinbib,showpacs]{revtex4-1}
\usepackage{graphicx,latexsym}

\def\bea{\begin{eqnarray}}
\def\eea{\end{eqnarray}}
\def\be{\begin{equation}}
\def\ee{\end{equation}}

\newcommand{\Pminus}{{\cal P}^-}

\begin{document}

\title{Basis of symmetric polynomials \\
for many-boson light-front wave functions}

\author{Sophia S. Chabysheva}
\author{John R. Hiller}
\affiliation{Department of Physics \\
University of Minnesota-Duluth \\
Duluth, Minnesota 55812}

\date{\today}

\begin{abstract}
We provide an algorithm for the construction of orthonormal
multivariate polynomials that are symmetric with respect to
the interchange of any two coordinates on the unit hypercube
and are constrained to the hyperplane where the sum of the
coordinates is one.  These polynomials form a basis for
the expansion of bosonic light-front momentum-space wave
functions, as functions of longitudinal momentum, where
momentum conservation guarantees that the fractions are
on the interval $[0,1]$ and sum to one.  This generalizes
earlier work on three-boson wave functions to wave functions
for arbitrarily many identical bosons.  A simple application
in two-dimensional $\phi^4$ theory illustrates the use
of these polynomials.
\end{abstract}

%

\maketitle

\section{Introduction}
\label{sec:introduction}

To solve a quantum field theory nonperturbatively, numerical techniques
are usually required.  The most commonly used technique is lattice gauge
theory~\cite{lattice}; however, this approach is Euclidean and lacks direct contact
with wave functions.  Without wave functions in a Minkowski metric,
some physical observables can be difficult if not impossible to 
calculate.  Access to wave functions provides for a much more direct
approach.  The technique of Dyson-Schwinger equations~\cite{DSE}
rectifies this situation somewhat, but remains Euclidean and requires
models for higher vertex functions.  An alternative that can provide
wave functions in Minkowski (momentum) space is the light-front
Hamiltonian approach~\cite{DLCQreview,LFwhitepaper,Vary}.

In light-front quantization, the state of a system is found as an eigenstate
of the Hamiltonians $\Pminus\equiv {\cal P}^0-{\cal P}^z$, 
${\cal P}^+\equiv {\cal P}^0+{\cal P}^z$, and 
$\vec{\cal P}_\perp\equiv ({\cal P}^x,{\cal P}^y)$.  Here ${\cal P}^0$
is the equal-time Hamiltonian operator, and $\vec{\cal P}=({\cal P}^x,{\cal P}^y,{\cal P}^z)$
is the equal-time momentum operator.  The light-front Hamiltonian $\Pminus$
evolves a system in light-front time $x^+\equiv t+z$; the momentum operator
${\cal P}^+$ translates a system in $x^-\equiv t-z$.  These are the
light-front coordinates of Dirac~\cite{Dirac}.  The stationary eigenstates
of $\Pminus$ can be expanded in a Fock basis consisting of eigenstates of
${\cal P}^+$ and $\vec{\cal P}_\perp$ and of the particle-number operator; 
only $\Pminus$ contains terms that change constituents and mix different
Fock states.  The coefficients
of the Fock-state expansion are the light-front momentum-space wave 
functions.

In the longitudinal (plus) direction, the light-front momentum of the
$i$th constituent $p_i^+=E_i+p_i^z$ is always positive.  We can then 
define longitudinal momentum fractions $x_i=p_i^+/P^+$, relative
to the total momentum $P^+$.  The wave functions are boost-invariant
functions of these momentum fractions.\footnote{This boost invariance
is one of the advantages of light-front quantization~\protect\cite{DLCQreview}.  
Another is that the Fock-state expansion itself is well defined; the positivity
of the longitudinal momentum prevents spurious vacuum contributions.}
Momentum conservation requires that these
fractions be on the interval $[0,1]$ and that the fractions sum to one.

In order that the Fock expansion be an eigenstate of $\Pminus$, the wave
functions must satisfy a system of integral equations.  One way to solve
such a system is to expand the wave functions in a truncated set of
basis functions and solve the resulting matrix eigenvalue problem for
the coefficients of the basis functions.

If a Fock state consists of identical bosons, the wave function
must be symmetric in its arguments.  If such a wave function is to be
expanded in a basis, the basis functions should also be symmetric.
If the momentum values are unconstrained, this is relatively straightforward,
but here the longitudinal momenta {\em are} constrained.  Therefore, 
for the dependence on longitudinal momenta,
we need a set of basis functions that satisfy the symmetry requirement
and the constraint.  For the case of two bosons, this is quite simple,
because there is only one independent variable.  For three bosons,
the analysis is somewhat complex; details can be found 
elsewhere~\cite{SymPolys}.  Here we study the general case, wave functions
for an arbitrary number of identical bosons.

Because the momentum fractions are limited to the interval $[0,1]$ and
constrained to sum to one, the wave function for $N$ bosons is defined
on an $(N-1)$-dimensional hyperplane within an $N$-dimensional hypercube.
Thus, multivariate polynomials with the correct symmetry, combined with
weight functions that control endpoint behavior, can be a convenient
choice for basis functions.

The construction of multivariate symmetric polynomials on the $N$-dimensional
hypercube is straightforward.  The difficulty comes from the fact that, when
restricted to the hyperplane defined by $\sum_i^N x_i=1$, most of the polynomials
of a given order are no longer linearly independent.  For example, consider 
the case of $N=4$.  As is true for any $N$, there is no linear polynomial,
because the only symmetric form on the hypercube, $\sum_i^N x_i$, is identically
equal to one on the hyperplane.  For second order polynomials, the two choices
on the hypercube, $\sum_i^4 x_i^2$ and $\sum_{i\leq j}^4 x_i x_j$, differ only
by multiplicative and additive constants on the hyperplane.  Similarly,
the three possible third-order polynomials on the hypercube are linearly related
on the hyperplane.  At order four, however, we find two linearly independent
polynomials on the hyperplane, reduced from five on the hypercube.  These can
be chosen to be the square of the second-order polynomial and the polynomial
$\prod_i^4 x_i$; any other symmetric polynomial of order four can be 
written as a linear combination of these two, plus lower-order polynomials.

Although it is possible, in principle, to continue working out polynomials
order by order for fixed $N$, as in the $N=4$ example above, this is not very
practical.  Instead we have developed an algorithm by which the structure
of these polynomials is known for arbitrary order and for an arbitrary number
$N$ of bosons.  Orthonormality, with respect to any appropriate weight
function, is then constructed order by order and polynomial by polynomial, 
with use of the Gramm-Schmidt process, which is easily automated.

The algorithm is derived and described in Sec.~\ref{sec:algorithm}.  A
simple application to two-dimensional $\phi^4$ theory, to illustrate the 
use of these basis functions, is discussed in Sec.~\ref{sec:application};
this includes a comparison with results obtained with the frequently
used method of discrete light-cone quantization (DLCQ)~\cite{PauliBrodsky}.  
A short summary is given in Sec.~\ref{sec:summary}.

\section{Algorithm}
\label{sec:algorithm}

We wish to construct linearly independent polynomials in $N$
variables $x_i$ that are fully symmetric on the unit hypercube
and restricted to the hyperplane defined by $\sum_i^N x_i=1$.
Those symmetric on the hypercube are easily constructed, as
\be \label{eq:tildePnN}
\tilde P_n^{(N)}=x_1^{\tilde n_1}x_x^{\tilde n_2}\cdots x_N^{\tilde n_N}+\mbox{permutations},
\ee
with $n=\tilde n_1+\tilde n_2+\cdots+\tilde n_N$ the order of the polynomial.
To make the polynomial unique, the individual powers are
restricted by the inequalities
\be \label{eq:constraint}
\tilde n_1\leq \tilde n_2\leq \cdots \leq \tilde n_N.
\ee
Unfortunately, these polynomials are not linearly independent
on the hyperplane $\sum_i^N x_i=1$, as discussed in the
Introduction.  

Consider, however, a different construction.  Given a set of
$N$ polynomials $C_m$ on the hypercube, where $C_m$ is of order $m$,
a polynomial of order $n$ can be built as
\be
P_n^{(N)}=C_1^{n_1}C_2^{n_2}\cdots C_N^{n_N},
\ee
with $n=n_1+2n_2+3n_3+\cdots+Nn_N$.  These polynomials have the
distinct advantage that the restriction to the hyperplane is
obvious; because $C_1=\sum_i x_i$ becomes 1, the linearly independent set
is obtained by considering only those $P_n^{(N)}$ for which $n_1=0$,
provided the other $C_m$ satisfy some restrictions, discussed below.

It is not immediately obvious that this new set is the same size
as the first.  Because it is linearly independent, it can be no
larger, but it could be smaller.  To see what happens, we can
simply count polynomials in each basis on the hypercube. This 
generalizes the proof for $N=3$, given in the Appendix of \cite{SymPolys}.

For the first set, the $\tilde P_n^{(N)}$, we have the following 
number of polynomials of order $n$:
\be
\tilde S_n^{(N)}=\sum_{\tilde n_1=0}^{[n/N]}\sum_{\tilde n_2=\tilde n_1}^{[(n-\tilde n_1)/(N-1)]}
  \cdots \sum_{\tilde n_i=\tilde n_{i-1}}^{[(n-\sum_{i'=1}^{i-1}\tilde n_{i'})/(N-i+1)]}
  \cdots \sum_{\tilde n_{N-1}=\tilde n_{N-2}}^{[(n-\sum_{i'}^{N-2}\tilde n_{i'})/2]}1,
\ee
with $[x]$ being the integer part of $x$.  The value of $\tilde n_N$ is
fixed at $n-\sum_i^{N-1}\tilde n_i$.  The lower limits on the sums
are determined by the constraint (\ref{eq:constraint}).  The upper limit
on $\tilde n_1$ can be no higher than $[n/N]$, because the other $N-1$ indices
must start at this upper limit and all together they must sum to $n$;
this would be impossible if $\tilde n_1$ went beyond $n/N$.  Similarly,
for $\tilde n_2$, the upper limit must be the available total of $n-n_1$
divided among the remaining $N-1$ indices.  Continuing in this fashion,
we determine all the upper bounds on the sums.\footnote{Notice that there 
is a typographical error in Eq.~(A2) of \protect\cite{SymPolys};
the upper limit of the first sum should be $[N/3]$.}  

For the second set of polynomials, the number of order $n$ on the 
hypercube is
\be
S_n^{(N)}=\sum_{n_N=0}^{[n/N]}\sum_{n_{N-1}=0}^{[(n-Nn_N)/(N-1)]}
  \cdots \sum_{n_i=0}^{[(n-\sum_{i'=i+1}^N i' n_{i'})/i]}
  \cdots \sum_{n_2=0}^{[(n-\sum_{i'=3}^N i' n_{i'})/2]} 1,
\ee
with $n_1=n-\sum_{i=2}^N i n_i$.  The upper bounds are determined by
the portion of the total order $n$ that can be assigned to a
particular polynomial $C_m$.  In general, this is at most $[n/m]$.
However, if other $C_m$ factors have already been assigned some
contribution to the total order, this contribution must first be subtracted
from $n$ before the division by $m$; the upper bound on a particular sum
takes this into account by subtracting from $n$ the appropriate contribution
already made in the sums to the left.

To show that the two counts $\tilde S_n^{(N)}$ and $S_n^{(N)}$ are the
same, introduce to $S_n^{(N)}$ the following change in summation indices:
\be
n_i=\left\{\begin{array}{ll} \tilde n_{N-i+1}-\tilde n_{N-i}, & i<N \\
                              \tilde n_1, & i=N. \end{array} \right.
\ee
The sum over $n_i$ becomes
\be
\sum_{n_i=0}^{[(n-\sum_{i'=i+1}^N i' n_{i'})/i]}
= \sum_{\tilde n_{N-i+1}=\tilde n_{N-i}}
    ^{[(n-\sum_{i'=i+1}^N i' (\tilde n_{N-i'+1}-\tilde n_{N-i'}))/i]+\tilde n_{N-i}},
\ee
with the understanding that $\tilde n_0=0$.
The upper limit can be simplified by taking advantage of cancellations.  For example,
the last two terms in the sum are 
$(N-1)(\tilde n_2-\tilde n_1)+N \tilde n_1=\tilde n_1+(N-1)\tilde n_2$.
The result is 
\be
[(n-\sum_{i'=i+1}^N i' (\tilde n_{N-i'+1}-\tilde n_{N-i'}))/i]+\tilde n_{N-i}
=[(n-\sum_{i'=1}^{N-i}\tilde n_{i'})/i-\tilde n_{N-i}]+\tilde n_{N-i}
=[(n-\sum_{i'=1}^{N-i}\tilde n_{i'})/i].
\ee
As a final step, we use this reduction and replace $i$ by $N-i+1$, to obtain
\be
\sum_{n_i=0}^{[(n-\sum_{i'=i+1}^N i' n_{i'})/i]}
\rightarrow \sum_{\tilde n_i=\tilde n_{i-1}}^{[n-\sum_{i'=1}^{i-1}\tilde n_i)/(N-i+1)]}.
\ee
Given this reduction of the individual sums, the count $S_n^{(N)}$ takes the
same form as $\tilde S_n^{(N)}$.  Thus, the number of linearly independent
polynomials on the hypercube is the same for the two forms $P_n^{(N)}$ and
$\tilde P_n^{(N)}$.  The restriction to the hyperplane then selects the 
subset of the $P_n^{(N)}$ with $n_1=0$.

The structure of symmetric polynomials in $N$ variables on the hyperplane 
can now be written as linear combinations of the factorizations
\be
P_{ni}^{(N)}=C_2^{n_2} C_3^{n_3}\cdots C_N^{n_N},
\ee
with the indices restricted by $n=\sum_j j n_j$.  The second subscript
on $P_{ni}^{(N)}$ differentiates between different linearly independent
polynomials of the same order $n$.  For $n\geq2$ there is always at least
one such polynomial.  For $n\geq4$, there can be more than one, depending
on the dimension $N$ of the hypercube.

One caveat is that the factorization of $P_{ni}^{(N)}$ assumes that none
of the $C_m$ can be written as a product of lower-order polynomials.
Such products are already included directly in the factorization.  For
example, a choice of $C_4=C_2^2$ would mean that the seventh-order
polynomial $C_3C_4$ is the same as the polynomial $C_2^2C_3$.  This
must be avoided by a proper choice of the set $\{C_m\}$; otherwise,
the counting argument is not valid, because it assumes linear independence
of the different factorizations.

A choice of the $C_m$ that maintains the linear independence, though
probably not unique, is to always write $C_m$ as a product of the 
lowest order monomials available.  They take the form of the 
$\tilde P_n^{(N)}$, as given in (\ref{eq:tildePnN}),
with all indices $\tilde n_i$ equal to zero or one and,
of course, summing to $n$.  For example, $C_N$ would be
$x_1x_2\cdots x_N$ and $C_{N-1}$ would be 
$x_2 x_3\cdots x_N+\mbox{permutations}$.  When restricted
to the hyperplane, $x_N$ is replaced by $1-\sum_i^{N-1} x_i$;
this generates terms that are no more than quadratic in 
any individual $x_i$, 
not only for $C_2$ but for all the $C_m$.  Any product of the
$C_m$ will contain higher powers of $x_i$ and will therefore
be automatically linearly independent of any individual $C_m$.

All that remains to complete the algorithm is to specify the
orthonormalization.  This is done by the standard Gramm-Schmidt 
construction, relative to a chosen inner product.  Given
a positive weight function $w(x_i)$ on the hyperplane,
the orthonormal combinations $O_{ni}^{(N)}$ of the basis
polynomials $P_{ni}^{(N)}$ are chosen to satisfy
\be \label{eq:innerproduct}
\int_0^1 dx_1\int_0^{1-x_1}dx_2 \cdots \int_0^{1-\sum_j^{N-2} x_j} dx_{N-1}
w(x_i) O_{n'i'}^{(N)}(x_i) O_{ni}^{(N)}(x_i) =\delta_{n'n}\delta_{i'i}.
\ee
We would then naturally choose basis functions for $N$-boson
wave functions as
\be
f_{ni}^{(N)}(x_i)=\sqrt{w(x_i)}O_{ni}^{N}(x_i).
\ee
The choice of weight function is driven by the particular
application and can be used to incorporate endpoint 
behavior in the basis functions.  For $N=2$ and a weight
function of the form $x_1^\alpha x_2^\beta$, the orthonormal
polynomials generated by the given algorithm are proportional
to the even-order Jacobi polynomials $P_n^{(\alpha,\beta)}$, 
transformed from $[-1,1]$ to $[0,1]$.

\section{Application}
\label{sec:application}

As an example of how these polynomials can be used, we consider 
two-dimensional $\phi^4$ theory.  The light-front Hamiltonian is
\be \Pminus=\Pminus_{11}+\Pminus_{22}+\Pminus_{13}+\Pminus_{31},
\ee
with
\bea \label{eq:Pminus11}
\Pminus_{11}&=&\int dp^+ \frac{\mu^2}{p^+} a^\dagger(p^+)a(p^+),  \\
\label{eq:Pminus22}
\Pminus_{22}&=&\frac{\lambda}{4}\int\frac{dp_1^+ dp_2^+}{4\pi\sqrt{p_1^+p_2^+}}
       \int\frac{dp_1^{\prime +}dp_2^{\prime +}}{\sqrt{p_1^{\prime +} p_2^{\prime +}}} 
       \delta(p_1^+ + p_2^+-p_1^{\prime +}-p_2^{\prime +}) \\
 && \rule{2in}{0mm} \times a^\dagger(p_1^+) a^\dagger(p_2^+) a(p_1^{\prime +}) a(p_2^{\prime +}),
   \nonumber \\
\label{eq:Pminus13}
\Pminus_{13}&=&\frac{\lambda}{6}\int \frac{dp_1^+dp_2^+dp_3^+}
                              {4\pi \sqrt{p_1^+p_2^+p_3^+(p_1^++p_2^++p_3^+)}} 
     a^\dagger(p_1^++p_2^++p_3^+)a(p_1^+)a(p_2^+)a(p_3^+), \\
\label{eq:Pminus31}
\Pminus_{31}&=&\frac{\lambda}{6}\int \frac{dp_1^+dp_2^+dp_3^+}
                              {4\pi \sqrt{p_1^+p_2^+p_3^+(p_1^++p_2^++p_3^+)}} 
      a^\dagger(p_1^+)a^\dagger(p_2^+)a^\dagger(p_3^+)a(p_1^++p_2^++p_3^+).
\eea
The boson creation and annihilation operators satisfy the commutation relation
\be
[a(p^+),a^\dagger(p^{\prime +})]=\delta(p^+-p^{\prime +}).
\ee
The creation operators can be used to construct Fock states
\be
|x_iP^+;P^+,n\rangle=\frac{1}{\sqrt{n!}}\prod_{i=1}^n a^\dagger(x_iP^+)|0\rangle
\ee
from the Fock vacuum $|0\rangle$.  An eigenstate of
$\Pminus$ can be written as an expansion in these Fock states
\be
|\psi(P^+)\rangle=\sum_n (P^+)^{(n-1)/2}
    \int \left(\prod_{i=1}^{n-1} dx_i\right) \psi_n(x_1,...,x_n) |x_iP^+;P^+,n\rangle,
\ee
with $\psi_n$ the $n$-boson wave function.

The terms of the Hamiltonian change particle number by zero or two,
and, therefore, eigenstates can be classified as having an even or 
odd number of constituents.  In \cite{SymPolys} we considered the
odd case, in order to have a Fock state with three bosons.  Here,
to have a more general application, we consider the even case
and truncate the basis to include only two and four-boson Fock
states.  To simplify the analysis, we also truncate the Hamiltonian
to discard the two-two scattering term $\Pminus_{22}$ in the four-boson sector;
this allows the system of equations for the wave functions to
be reduced to a single two-body equation, where a high-resolution
calculation can be easily made, for the purpose of comparison.

The light-front eigenvalue problem~\cite{DLCQreview}
\be
\Pminus|\psi(P^+)\rangle=\frac{M^2}{P^+}|\psi(P^+)\rangle \;\;
\mbox{and} \;\; 
{\cal P}^+|\psi(P^+)\rangle=P^+|\psi(P^+)\rangle
\ee
yields the following coupled integral equations for the
two-boson and four-boson wave functions:
\bea \label{eq:twoboson}
M^2\psi_2&=&\left(\frac{\mu^2}{x_1}+\frac{\mu^2}{x_2}\right)\psi_2
  +\frac12 \frac{\lambda}{4\pi}\frac{1}{\sqrt{x_1x_2}}\int_0^1 \frac{dx'_1}{\sqrt{x'_1x'_2}}\psi_2(x'_1,x'_2) \\
  &&+\frac{1}{\sqrt{3}} \frac{\lambda}{4\pi}\int_0^{x_1} dx'_1\int_0^{x_1-x'_1}dx'_2
     \left(\frac{\psi_4(x'_1,x'_2,x_1-x'_1-x'_2,x_2)}{\sqrt{x_1x'_1x'_2(x_1-x'_1-x'_2)}}
                       +(x_1\leftrightarrow x_2)\right), \nonumber
\eea
\be \label{eq:fourboson}
M^2\psi_4=\sum_{i=1}^4\frac{\mu^2}{x_i}\psi_4
     +\frac{1}{2\sqrt{3}}\frac{\lambda}{4\pi}\left(\frac{\psi_2(x_1+x_2+x_3,x_4)}{\sqrt{x_1x_2x_3(x_1+x_2+x_3)}}
        +(x_1\leftrightarrow x_4)+(x_2\leftrightarrow x_4)+(x_3\leftrightarrow x_4)\right).
\ee
The second equation can be solved explicitly for $\psi_4$.  Substitution
into the first equation provides a single, two-boson equation,
\bea \label{eq:reduced}
M^2\psi_2&=&\left(\frac{\mu^2}{x_1}+\frac{\mu^2}{x_2}\right)\psi_2
+\frac12 \frac{\lambda}{4\pi}\frac{1}{\sqrt{x_1x_2}}\int_0^1 \frac{dx'_1}{\sqrt{x'_1x'_2}}\psi_2(x'_1,x'_2) \\
  && +\frac16\left(\frac{\lambda}{4\pi}\right)^2\int_0^{x_1} \frac{dx'_1}{x'_1}\int_0^{x_1-x'_1}\frac{dx'_2}{x'_2}
  \left[\left(\frac{1}{M^2-\frac{\mu^2}{x'_1}-\frac{\mu^2}{x'_2}-\frac{\mu^2}{x_1-x'_1-x'_2}-\frac{\mu^2}{x_2}}\right)
    \right.   \nonumber \\
 && \left.  \times
  \left(\frac{\psi_2(x_1,x_2)}{x_1(x_1-x'_1-x'_2)}
      +3\frac{\psi_2(x_1-x'_1-x'_2,x'_1+x'_2+x_2)}{\sqrt{x_1x_2(x_1-x'_1-x'_2)(x'_1+x'_2+x_2)}}\right)
        +(x_1\leftrightarrow x_2)\right].  \nonumber
\eea
In each of these equations it is to be understood that the second momentum fraction
for a two-boson system and the fourth momentum fraction for a four-boson system
is not truly independent; the sum of momentum fractions in a wave function must be one.

To use the symmetric orthonormal polynomials $O_{ni}^{(N)}$ to solve the system of 
equations, we approximate the wave functions by truncated sums
\be
\psi_2(x_1,x_2)=\sqrt{x_1x_2} \sum_n^K a_n^{(2)} O_n^{(2)}(x_1)
\ee
and
\be
\psi_4(x_1,x_2,x_3,x_4)=\sqrt{x_1x_2x_3x_4} \sum_{ni}^K a_{ni}^{(4)}  O_{ni}^{(4)}(x_1,x_2,x_3).
\ee
The truncation at $n=K$ is a truncation of the range of polynomial
orders to a maximum of $K$.  Although the truncations can be tuned
separately in the different Fock sectors, to optimize a calculation,
we do not do that here.  Also, so that the $N=2$ polynomials change
as the truncation $K$ is relaxed, we take $K$ to be even and
increment in steps of 2 when studying convergence, there being no 
odd-order symmetric polynomials for two bosons.

\begin{figure}[ht]
\vspace{0.3in}
\centerline{\includegraphics[width=11cm]{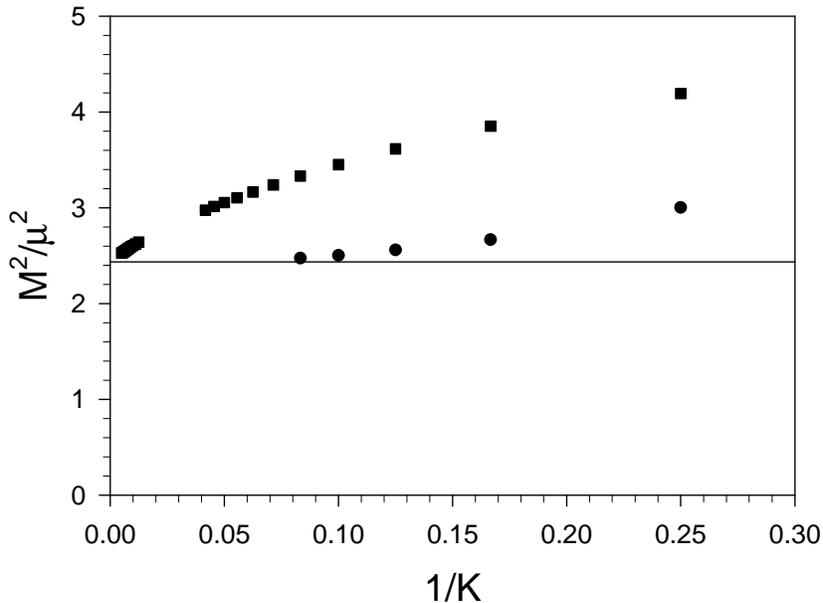}}
\caption{\label{fig:M2vsK} Comparison of convergence rates for the
fully symmetric polynomial basis (filled circles) and DLCQ (filled squares).  
The dimensionless eigenmass $M^2/\mu^2$ is plotted versus $1/K$, the reciprocal 
of the basis order and of the DLCQ resolution, for the case where 
the coupling is $\lambda=4\pi\mu^2$.  The lower resolution DLCQ results
cover a range up to $K=24$; the higher resolution results, a range
of 80 to 200.  The higher resolution points are used for an extrapolation
to $K=\infty$, and the horizontal line is at the value of that limit.
}
\end{figure}

On substitution of the polynomial expansions, the coupled integral
equations (\ref{eq:twoboson}) and (\ref{eq:fourboson}) become
\bea
\frac{M^2}{\mu^2}a_m^{(2)}&=&A^{(2)}_{mn}a_n^{(2)}+\frac12\frac{\lambda}{4\pi\mu^2}B_m B_n a_n^{(2)}
   +\frac{2}{\sqrt{3}}\frac{\lambda}{4\pi\mu^2}C_{m,ni}a_{ni}^{(4)}, \\
\frac{M^2}{\mu^2}a_{mj}^{(4)}&=&A^{(4)}_{mj,ni}a_{ni}^{(4)}
      +\frac{2}{\sqrt{3}}\frac{\lambda}{4\pi\mu^2}C_{n,mj}a_n^{(2)},
\eea
where repeated indices are summed and the matrices are
\bea
A^{(2)}_{mn}&=&2\int_0^1 dx_1 (1-x_1) O_m^{(2)}(x_1)O_n^{(2)}(x_1), \\
A^{(4)}_{mj,ni}&=&4\int_0^1 dx_1 x_2 x_3 (1-x_1-x_2-x_3)   
                    O_{mj}^{(4)}(x_1,x_2,x_3)O_{ni}^{(4)}(x_1,x_2,x_3), \\
B_n&=&\int_0^1 dx_1 O_n^{(2)}(x_1), \\
C_{m,ni}&=&\int_0^1 dx_1 \int_0^{1-x_1}dx_2\int_0^{1-x_1-x_2}dx_3 (1-x_1-x_2-x_3)
              O_m^{(2)}(x_1+x_2+x_3) O_{ni}^{(4)}(x_1,x_2,x_3). \nonumber \\
\eea
These algebraic equations define a matrix eigenvalue problem that is
readily solved, once the individual overlap integrals have been done
to compute the matrix elements.  The results for a series of truncations
at a fixed coupling of $\lambda=4\pi\mu^2$ are shown in Fig.~\ref{fig:M2vsK}.
%

For comparison, we consider a DLCQ approximation~\cite{PauliBrodsky},
which is, in the context of these integral equations, equivalent to
a trapezoidal approximation to the integrations with a step size
that is a reciprocal of an integer $K$ and where the endpoint
contributions are assumed to be zero.  In DLCQ, $K$ is called
the harmonic resolution~\cite{PauliBrodsky} or just `resolution.'
The individual momentum fractions are resolved into multiples
of $1/K$.  The neglect of the endpoints is the exclusion
of zero modes, modes of zero longitudinal momentum.  This
is a standard approximation in DLCQ, but it does delay 
convergence, because the wave functions go to zero
slowly enough that the integrations do have endpoint
contributions.  Thus the neglect of zero modes induces
errors of order $1/K$; however, inclusion of them within
the full many-body DLCQ approach is nontrivial~\cite{DLCQzeromodes}.

The DLCQ approximation can be applied to either the coupled set or,
equivalently, to the reduced two-boson equation, (\ref{eq:reduced}).
We again obtain a matrix eigenvalue problem:
\be
\frac{M^2}{\mu^2}\psi_{2m}=\left(\frac{K}{m}+\frac{K}{K-m}\right)\psi_{2m}
   +\frac12\frac{\lambda}{4\pi\mu^2}\tilde B_{mn}\psi_{2n}
  +\frac12\left(\frac{\lambda}{4\pi\mu^2}\right)^2\tilde C_{mn}\psi_{2n}
  +\frac16\left(\frac{\lambda}{4\pi\mu^2}\right)^2\tilde\Delta_m \psi_{2m},
\ee
where $\psi_{2m}\equiv \psi_2(m/K,(K-m)/K)$ and
\bea
\tilde B_{mn}&=&\frac{K}{\sqrt{m(K-m)n(K-n)}}, \\
\tilde C_{mn}&=&\frac{1}{K^2}\sum_{n_1=1}^{m-n-1}
     \frac{1}{\frac{M^2}{\mu^2}-\frac{K}{n_1}-\frac{K}{m-n_1-n}-\frac{K}{n}-\frac{K}{K-m}}
     \frac{1}
        {\frac{n_1}{K}\frac{m-n_1-n}{K}\sqrt{\frac{m}{K}\frac{K-m}{K}\frac{n}{K}\frac{K-n}{K}}} \\
      &&  +(m\leftrightarrow K-m,\;n\leftrightarrow K-n),  \nonumber \\
\tilde\Delta_m&=&\frac{1}{K^2}\sum_{n_1=1}^{m-1}\sum_{n_2=1}^{m-n_1-1}
   \frac{1}{\frac{M^2}{\mu^2}-\frac{K}{n_1}-\frac{K}{n_2}-\frac{K}{m-n_1-n_2}-\frac{K}{K-m}}
   \frac{1}{\frac{n_1}{K}\frac{n_2}{K}\frac{m-n_1-n_2}{K}\frac{m}{K}}
       +(m\leftrightarrow K-m). \nonumber \\  
\eea
However, the matrices themselves depend upon the eigenvalue.  This (expected) complication
for the reduced equation is easily managed, by iteration from a guess for the eigenmass.
For a given value of $M^2$ used in constructing the right-hand side, the lowest eigenvalue of
the matrix can be computed and compared with the chosen value.  If they do not
agree, a new estimate of the eigenmass can be formed and the process repeated.
We used the Muller algorithm~\cite{Muller} to guide the iterations; this improves on 
the more common secant algorithm with use of a quadratic, rather than linear, fit.

Some results are shown in Fig.~\ref{fig:M2vsK}.  The low resolution DLCQ results are
far from convergence.  The high resolution results are quite close and easily 
extrapolated.  However, these resolutions, from $K=80$ to 200, are well beyond what
can be used in practice for a many-body DLCQ calculation; the state of the art for 
$\phi^4$ theory has been extended to $K=72$ on massively parallel machines~\cite{VaryKlimit}.
The primary reason for DLCQ's slow convergence is the poor endpoint behavior.
The expansions in terms of symmetric polynomials have the freedom to adjust
the endpoint behavior in a very straightforward fashion.  Convergence is then
much more rapid.

\section{Summary}
\label{sec:summary}

We have derived an algorithm for the construction of fully symmetric
orthonormal multivariate polynomials for the representation of
the longitudinal momentum dependence of light-front wave functions
for arbitrarily many bosons.  The orthonormalization is carried out
by the standard Gramm-Schmidt process.  This process is applied
to the symmetric polynomials for $N$ bosons,
obtained by considering all possible
factorizations of the form $\prod_{m=2}^N C_m^{n_m}$, where the
order $n$ of the polynomial can be decomposed as $n=\sum_m^N m\,n_m$
and $C_m$ is an order-$m$ polynomial with $2<=m<=N$.  The orthonormalization
can be done relative to an inner product, such as (\ref{eq:innerproduct}),
with a weight function chosen to optimize the utility of the polynomials.
In particular, the weight function can be coordinated with the expected
endpoint behavior of the wave functions to be represented.
The example shown here, in Sec.~\ref{sec:application}, illustrates
the dramatically improved convergence, compared to the DLCQ method.

The original motivation in seeking these polynomials was in 
applications to equations obtained in the light-front 
coupled-cluster (LFCC) method~\cite{LFCC,LFCCphi4}.  There the
function of interest, to be expanded using these polynomials
as a basis, is not a wave function but instead a vertex-like
function that controls the operator that generates wave functions.
This is done to avoid making a Fock-space truncation.  However,
the linearized version of the LFCC equations is equivalent to
the Fock-space wave-function equations considered here.

The LFCC method will never use more than a small range of
the boson multiplicity $N$; for the current work on applications
to $\phi^4$ theory~\cite{LFCCphi4}, $N=3$ and 4 are enough.
Where the generalization to arbitrary $N$ is important is
in direct applications to Fock-state expansions
for many-boson problems in light-front
quantization.  Specifically, the light-front many-body problem for
$\phi^4$ theory, which has been attempted only with DLCQ~\cite{VaryKlimit}, 
can now be attacked with polynomial expansions in each Fock
sector.  

Such expansions are superior to DLCQ in two 
respects.  One is the control of endpoint contributions,
and the other is sector by sector control of resolution.
The endpoint contributions are critical, not only for rapid
convergence but also for computation of the vacuum
expectation value for the $\phi$ field when degenerate
odd and even eigenstates are mixed.  

The control of resolution in each sector is important for 
shifting computational resources to where they are most needed.
In DLCQ, the number of discrete Fock states in each
sector is fixed once the resolution is chosen, and this
number is quite large for sectors with boson numbers
near $K/2$, even though these sectors may not be 
particularly important for the calculation.  With
the polynomial expansion, the number and order of
polynomials used in any Fock sector can be set
individually, to place higher resolution in the
sectors found to be most important for a given
calculation.

\acknowledgments
This work was supported in part by the Minnesota Supercomputing Institute.




\end{document}